\documentclass[%
reprint,
superscriptaddress,
showpacs,
preprintnumbers,
 bibnotes,
 amsmath,amssymb,
 aps,
prl
]{revtex4-1}

\usepackage{graphicx}
\usepackage{dcolumn}
\usepackage{bm}
\usepackage{color}

\begin{document}


\title{Relationship among superconductivity, pseudogap, and high-energy magnetic fluctuations in a model high-${T_\mathrm{c}}$ superconductor from electronic Raman scattering}

\author{Yuan~Li}
\email[]{yuan.li@fkf.mpg.de}
\affiliation{Max Planck Institute for Solid State Research, D-70569 Stuttgart, Germany}
\affiliation{International Center for Quantum Materials, School of Physics, Peking University, Beijing 100871, China}
\author{M.~Le~Tacon}
\email[]{m.letacon@fkf.mpg.de}
\affiliation{Max Planck Institute for Solid State Research, D-70569 Stuttgart, Germany}
\author{M.~Bakr}
\altaffiliation[Present address: ]{Zentrum f\"ur Synchrotronstrahlung, TU Dortmund, D-44221 Dortmund, Germany}
\affiliation{Max Planck Institute for Solid State Research, D-70569 Stuttgart, Germany}
\author{D.~Terrade}
\author{D.~Manske}
\affiliation{Max Planck Institute for Solid State Research, D-70569 Stuttgart, Germany}
\author{R.~Hackl}
\affiliation{Walther Meissner Institute, Bavarian Academy of Sciences, D-85748 Garching, Germany}
\author{L.~Ji}
\affiliation{School of Physics and Astronomy, University of Minnesota, Minneapolis, Minnesota 55455, USA}
\author{M.~K.~Chan}
\affiliation{School of Physics and Astronomy, University of Minnesota, Minneapolis, Minnesota 55455, USA}
\author{N.~Bari\v{s}i\'{c}}
\affiliation{School of Physics and Astronomy, University of Minnesota, Minneapolis, Minnesota 55455, USA}
\author{X.~Zhao}
\affiliation{School of Physics and Astronomy, University of Minnesota, Minneapolis, Minnesota 55455, USA}
\affiliation{State Key Lab of Inorganic Synthesis and Preparative Chemistry, College of Chemistry, Jilin University, Changchun 130012, P.R. China}
\author{M.~Greven}
\affiliation{School of Physics and Astronomy, University of Minnesota, Minneapolis, Minnesota 55455, USA}
\author{B.~Keimer}
\affiliation{Max Planck Institute for Solid State Research, D-70569 Stuttgart, Germany}


\begin{abstract}
We use electronic Raman scattering to study the model single-layer cuprate superconductor HgBa$_2$CuO$_{4+\delta}$. In an overdoped sample, we observe a pronounced amplitude enhancement of a high-energy peak related to two-magnon excitations in insulating cuprates upon cooling below the critical temperature $T_\mathrm{c}$. This effect is accompanied by the appearance of the superconducting gap and a pairing peak above the gap in the Raman spectrum, and it can be understood as a consequence of feedback of the Cooper pairing interaction on the high-energy magnetic fluctuations. All of these effects occur already above $T_\mathrm{c}$ in two underdoped samples, demonstrating a related feedback mechanism associated with the pseudogap.
\end{abstract}

\maketitle


High-temperature superconductivity in the cuprates arises from doping antiferromagnetic (AF) insulators. This has motivated intense research on the role of AF fluctuations in the mechanism of superconductivity \cite{Scalapino1995,*AbanovAdvPhys2003,*MoriyaRepProgPhys2003}. Near the AF ordering wave vector and below the critical temperature $T_\mathrm{c}$, inelastic neutron scattering (INS) experiments have uncovered a pronounced spectral-weight redistribution of low-energy magnetic excitations into a ``resonance'' peak with energy 40-60 meV \cite{Sidis2004,*YuPRB2010}. The magnetic resonance appears generic to superconductors near an AF instability, including the cuprates \cite{Sidis2004,*YuPRB2010}, the heavy-fermion compounds \cite{SatoNature2001,*StockPRL2008}, and the iron-based superconductors \cite{ChristiansonNature2008,*LumsdenPRL2009,*InosovNatPhys2010}, and its energy scales with the superconducting gap \cite{YuNatPhys2009,*InosovPRB2011}. Based on these observations, and on related anomalies in fermionic spectral functions, the resonance has been attributed to a feedback effect of the Cooper pairing interaction on low-energy spin fluctuations that mediate pairing \cite{Eschrig2006}. However, the spectral weight of these low-energy fluctuations appears insufficient to explain the large $T_\mathrm{c}$ in the cuprates \cite{Eschrig2006}. Meanwhile, evidence from tunneling \cite{Pasupathy2008}, photoemission \cite{Kordyuk2006}, and optical \cite{Hwang2007,*vanHeumenPRB2009} spectroscopies has indicated contributions from high-energy excitations to the pairing interaction.

Recent research has begun to explore the origin of this high-energy contribution by extending the experimental study of magnetic excitations to higher energies. A strong magnetic response well above 100 meV has been found by INS in overdoped La$_{1.78}$Sr$_{0.22}$CuO$_4$ \cite{LipscombePRL2007} and by resonant inelastic $x$-ray scattering (RIXS) in various cuprates up to optimal doping \cite{LeTaconNatPhys2011}. These results demonstrate that high-energy fluctuations akin to magnons in the AF parent compounds are available as a possible resource for Cooper pairing deep in the superconducting regime of the phase diagram. However, it remains largely unknown whether this resource is actually utilized. In order to address this question, we have performed an accurate electronic Raman scattering (ERS) study of the model single-layer system HgBa$_2$CuO$_{4+\delta}$ (Hg1201) \cite{BarisicPRB2008}. Our results provide detailed information about the temperature evolution of the magnetic fluctuations that is difficult to obtain by INS and RIXS due to limited beam-time resources. With decreasing temperature, we observe an amplitude enhancement and an energy shift of a ``two-magnon'' peak attributable to high-energy magnetic fluctuations, which is accompanied by the opening of a gap and the appearance of a pairing peak above the gap. This effect occurs at $T_\mathrm{c}$ in an overdoped sample, and can hence be understood as a high-energy feedback effect analogous to the resonant mode observed by INS. In underdoped samples, we observe the same phenomena at temperatures well above $T_\mathrm{c}$. This suggests that a related feedback mechanism is operative in the pseudogap regime \cite{NormanAdvPhys2005}.

We studied three Hg1201 single crystals: strongly underdoped ($T_\mathrm{c}=77$~K, UD77), slightly underdoped ($T_\mathrm{c}=94$~K, UD94), and overdoped ($T_\mathrm{c}=90$~K, OV90), which have hole concentrations $p=0.11$, 0.14, and 0.19, respectively, according to the empirical equation \cite{TallonPRB1995} $T_\mathrm{c}/T_\mathrm{c}^\mathrm{max} = 1-82.6 (p-0.16)^2$ with $T_\mathrm{c}^\mathrm{max}=97$~K \cite{BarisicPRB2008}. The crystals were grown by a self-flux method \cite{ZhaoAdvMater2006}. Sharp transitions at $T_\mathrm{c}$,  a large diamagnetic signal below $T_\mathrm{c}$ in field-cooled measurements \cite{BarisicPRB2008}, and the observation of a long-range ordered magnetic vortex lattice in one of the samples (UD94) \cite{LiPRB2011} demonstrate the high quality of our samples. Hg1201 is nearly ideal for ERS experiments, because its simple tetragonal structure with only one CuO$_2$ plane per unit cell minimizes the number of Raman-active phonons and facilitates measurements in pure symmetry channels.

Our ERS measurements were performed in backscattering geometry using a JobinYvon LabRam 1800 single-grating spectrometer, equipped with two razor-edge filters to suppress the elastic line. Optically flat sample surfaces parallel to the CuO$_2$ planes were freshly polished and maintained in $<10^{-5}$ mbar vacuum for the measurements. The data presented here were obtained in the $B_{1g}$ geometry, which is sensitive to electronic excitations from the antinodal regions of reciprocal space \cite{DevereauxRMP2007}. To utilize resonant enhancement of electronic signals in this geometry \cite{LeTacon_PRB05}, we used the $\lambda=632.8$ nm line of a He-Ne laser for excitation. The 1.1 mW laser beam was focused onto a $\sim5\,\mu$m-diameter spot on the sample, limiting laser-heating to approximately 10 K. We present our data as the Raman susceptibility $\chi^{\prime\prime}$, which was derived by correcting the raw spectra for the spectrometer's efficiency, and dividing the intensities above a constant level $I_0$ by the Bose factor. $I_0$ was chosen such that $\chi^{\prime\prime}$ extrapolates to zero in the $\omega\rightarrow0$ limit.

\begin{figure}
\includegraphics[width=3.375in]{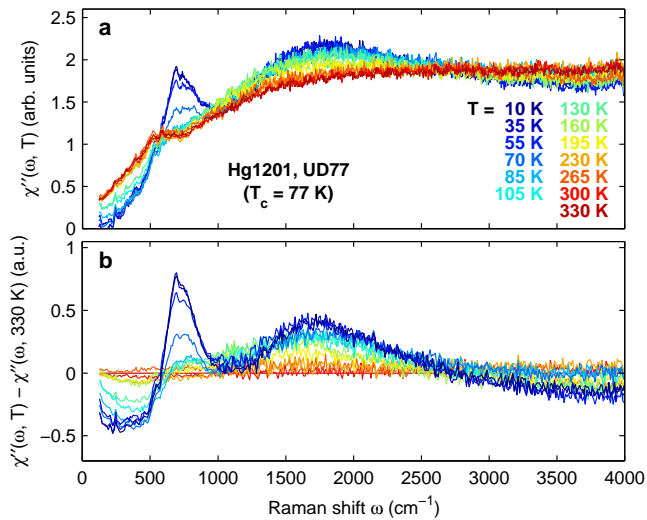}
\caption{\label{fig:one}
(a) Raman spectra for sample UD77. (b) Differential spectra relative to 330 K.
}
\end{figure}

\begin{figure}
\includegraphics[width=3.375in]{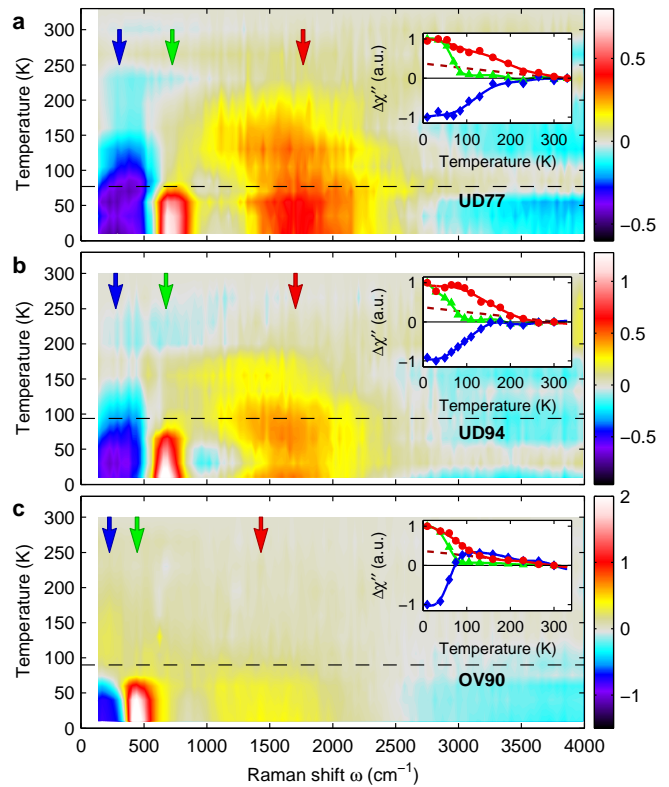}
\caption{\label{fig:two}
Main panels: Color plots of $\Delta\chi^{\prime\prime}$ relative to the highest temperatures. Dashed lines indicate $T_\mathrm{c}$.  Insets: $\Delta\chi^{\prime\prime}$ at energies indicated by the color-coded arrows, normalized to the lowest temperature. Solid curves are guides to the eye. Dashed lines (identical in all insets) describe the high-temperature behavior of the two-magnon amplitude in OV90.
}
\end{figure}

\begin{figure}
\includegraphics[width=3.375in]{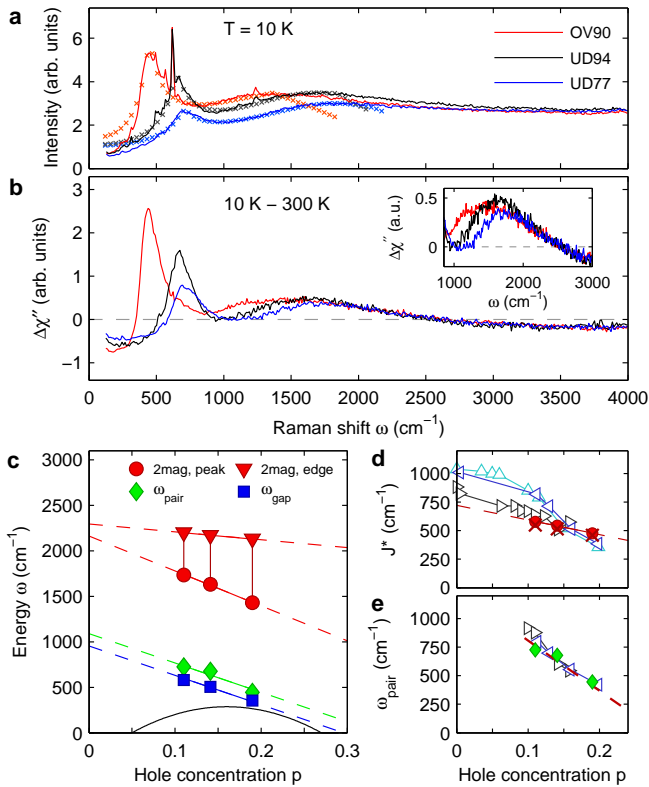}
\caption{\label{fig:four}
(a) Solid curves, raw ERS spectra at 10 K. Cross symbols, model calculations for $\omega$ up to $4J^*$. (b) $\Delta\chi^{\prime\prime}$ between 10 K and 300 K. Phonon peaks in UD94 and OV90 are removed prior to the subtraction. Inset shows an enlarged view of the two-magnon peak. (c) Characteristic energies in Hg1201 (see text). Dome-shaped curve, $4.28\,k_\mathrm{B}\,T_\mathrm{c}$. (d) Values of $J^*=\omega_\mathrm{2mag}^\mathrm{peak}/3$ (except for cross symbols which are from the calculations in (a)), and (e) $\omega_\mathrm{pair}$, in La$_{2-x}$Sr$_x$CuO$_4$ (upward triangles), YBa$_2$Cu$_3$O$_{6+\delta}$ (rightward triangles), Bi$_{2.1}$Sr$_{1.9}$Ca$_{1-x}$Y$_x$Cu$_2$O$_{8+\delta}$ (leftward triangles) \cite{SugaiPRB2003}, and Hg1201 (filled symbols, this work). Dashed line in (e) summarizes $\omega_\mathrm{pair}$ for Hg1201 reported in Ref.~\onlinecite{LeTaconNatPhys2006,*BlancPRB2010}.
}
\end{figure}

Figure~\ref{fig:one}a displays our results for UD77 over a wide energy ($\omega$) range from 120 to 4000~cm$^{-1}$ and for temperatures ($T$) between 10 and 330 K. The spectra show three key features, which we refer to using nomenclature consistent with the literature \cite{DevereauxRMP2007,GuyardPRB2008}: (1) the ``pseudogap'', which manifests itself as a depletion of spectral weight below 570 cm$^{-1}$, (2) the ``pairing peak'' centered at 725 cm$^{-1}$, and (3) the ``two-magnon peak'' at approximately 1700 cm$^{-1}$. The energy of feature (1) is consistent with the pseudogap observed by angle-resolved photoemission near the antinodes of the superconducting gap function at comparable doping levels \cite{LeeNature2007}. Feature (2) had long been associated with Cooper pair breaking \cite{DevereauxRMP2007}, but recent results have cast doubt on this interpretation \cite{MunnikesPRB2011}. Nonetheless, its temperature dependence (see below) indicates that it is directly related to superconductivity. The peak energy is consistent with the extrapolation of previous results for Hg1201 at higher doping levels \cite{LeTaconNatPhys2006,*BlancPRB2010} and results for other cuprates at similar doping levels \cite{KendzioraPRB1995,SugaiPRB2003,MunnikesPRB2011} (Fig.~\ref{fig:four}e). To the best of our knowledge, however, Fig.~\ref{fig:one} contains the clearest observation of the pairing peak for a doping level as low as UD77 ($p=0.11$). Feature (3) arises from high-energy electronic fluctuations that smoothly evolve with doping out of the two-magnon excitations in AF parent compounds \cite{TassiniPRB2008,SugaiPRB2003}. Although additional quantum phases and correlations may play some role \cite{VivekPRL2007,CapraraPRB2011}, the dominant character of these fluctuations thus appears to be closely related to high-energy magnons in the AF insulators.

We now discuss the evolution of the ERS spectra with temperature, which is best seen in the differential spectra $\Delta\chi^{\prime\prime}$ after subtracting the 330 K data (Fig.~\ref{fig:one}b). Figure~\ref{fig:two} displays $\Delta\chi^{\prime\prime}$ for all samples. The three key features are indicated by arrows color-coded with constant-energy plots in the insets. Our main finding pertains to the temperature dependence of the two-magnon peak (red arrow) and its correlation with the other features. We begin our discussion with the slightly overdoped sample, OV90. Upon cooling from 300 K, the two-magnon signal amplitude first increases linearly with decreasing $T$ (dashed line in Fig.~\ref{fig:two}c inset), indicating a slight reduction in thermal broadening. Then, near $T_\mathrm{c}$, the signal amplitude increases rapidly, in concert with the development of the gap (blue arrow) \footnote{A maximum of $\chi^{\prime\prime}$ in the energy range of the gap is found at intermediate temperature in OV90, consistent with previous results \cite{VenturiniPRL2002,GuyardPRB2008}. No $\omega/T$-scaling \cite{VarmaPRL1989} is found.} and the pairing peak (green arrow). The $T$-dependence of the two-magnon peak amplitude is in fact strikingly similar to that of the low-energy resonant mode observed by INS \cite{Sidis2004,*YuPRB2010}, suggesting a related interpretation as a feedback effect of Cooper pairing on the magnetic fluctuation spectrum. The observation of such a feedback for energies far above the superconducting gap (which is directly visible in the ERS spectra, Fig.~\ref{fig:two}c) is new and surprising \footnote{An increase of the two-magnon intensity with cooling has been reported \cite{RuebhausenPRB1997,RuebhausenPRL1999}, but the limited information on the low-energy features and on the temperature dependence makes it difficult to identify any correlation.}.

We now turn to the underdoped samples UD77 and UD94 (Fig.~\ref{fig:two}a-b), where the pseudogap opens up at a characteristic temperature $T_\mathrm{gap}$ well above $T_\mathrm{c}$ and evolves smoothly through $T_\mathrm{c}$. The same trend is observed for the anomaly in the $T$-dependent intensity of the two-magnon peak, which shifts to progressively higher temperatures with decreasing doping level. The highly accurate data on the two-magnon peaks in UD94 and UD77 also reveal a slight increase of its energy below $T_\mathrm{gap}$ (``banana shape'' in the color plots). This further confirms the correlation between these features and demonstrates that a feedback mechanism akin to the one observed in OD90 is also present in the pseudogap regime.

\begin{figure}
\includegraphics[width=3.375in]{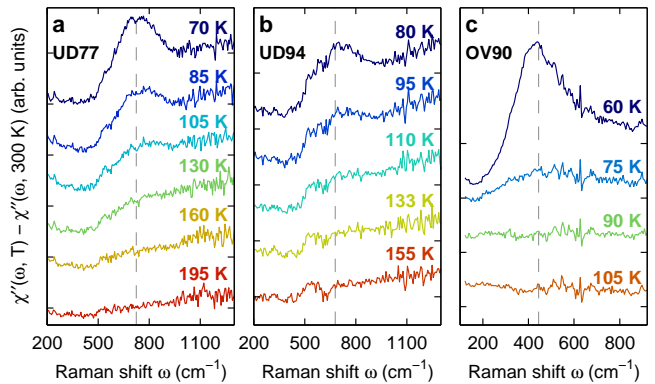}
\caption{\label{fig:three}
Differential spectra relative to 300 K near the pairing peak. Phonon peaks in UD94 and OV90 were removed prior to the subtraction. Data are offset for clarity. Dashed lines indicate the peak positions determined at 10 K.
}
\end{figure}

The pairing peak continues to exhibit a strong anomaly at $T_\mathrm{c}$ in the underdoped samples (Fig.~\ref{fig:two}a-b). However, close inspection of our data (Fig.~\ref{fig:three}) reveals remnant signals at the pairing peak energy up to 130 K in UD77 and 110 K in UD94. This has not been observed in previous ERS studies on underdoped cuprates \footnote{A doping-independent peak at 600 cm$^{-1}$ has been observed above $T_\mathrm{c}$ in Bi2212 \cite{BlumbergScience1997}, but subsequent studies suggested that the peak is phonon-related \cite{QuiltyPRB1998,*HewittPRB1999}.}, probably due to the peak's weak intensity in underdoped systems \cite{DevereauxRMP2007,SugaiPRB2003,LeTaconNatPhys2006,*BlancPRB2010} and/or the presence of impurities and strains \cite{MunnikesPRB2011}. These difficulties have been overcome in our study (Fig.~\ref{fig:one}).  The onset temperatures of the pairing peaks in UD77 and UD94 are well above $T_\mathrm{c}$ and not far from $T_\mathrm{gap}$, as can be seen from the tails of the green curves in the insets of Fig.~\ref{fig:two}a-b. In contrast, no extra intensity can be detected already at $T_\mathrm{c} = 90$ K in OV90 (Fig.~\ref{fig:three}c). Despite some quantitative differences in the onset temperatures of the three ERS features in UD77 and UD94 that presumably reflect their different energy scales, the correlation among their doping dependences is a very robust result.

In order to put the analysis of the doping dependence of the three spectral features on a quantitative footing, we have performed model calculations based on the $t$-$t'$-$J$ Hamiltonian, $H = H_{t,t'} + H_J$. We calculate the spectral response including both the pairing peak (following \cite{DevereauxPRB1995,*DevereauxPRB1996} using an {\it ab initio} tight-binding energy dispersion given in \cite{AndersenJPCS1995}) and the two-magnon peak. The intensity of the latter \cite{FleuryPR1968,CanaliPRB1992,RuebhausenPRB1997} is proportional to $\mbox{Im}\left[R(\omega)\left[1 + \left(1/Sz\alpha \right)R(\omega)\right]^{-1}\right]$ with
\begin{eqnarray}
R(\omega) = - 4\sum_{\bf k}f_k^2\,\frac{\omega_k + \Sigma({\bf k},\omega)}{\omega^2 - 4\left[\omega_k + \Sigma({\bf k},\omega)\right]^2}
\, . \label{eq:intensity}
\end{eqnarray}
Here, $f_k$ is the $B_\mathrm{1g}$ symmetry factor, $\Sigma({\bf k},\omega) = \Sigma^0 - i\Gamma$ is the self-energy of the one-magnon Green's function (treated as a phenomenological parameter), and $\omega_k = \frac{J^* S z}{\hbar}\sqrt{1 - \gamma_k^2}$ with $\gamma_k = \left[\cos(k_xa) + \cos(k_ya)\right]/2$ is the magnon dispersion. $\alpha = 1.158$ is a numerical constant \cite{CanaliPRB1992}, and $S=1/2$,
$z=4(1-p)$, and $a$ are the quantum number of spin, average number of nearest neighbors, and the in-plane lattice spacing, respectively. In general, our analysis of $H_J$ is valid up to the energy of undamped magnon excitations $\omega \simeq 4J^*$ where $J^*$ is an effective doping-dependent exchange parameter. Interference effects between $H_{t,t'}$ and $H_J$ are neglected. Additional phonon peaks in UD94 and OV90 (but not in UD77), possibly due to oxygen superstructures \cite{IzquierdoJPhysChemSolids2011}, are not considered.  Using AF interaction and gap parameters $J^*=548$, 516, and 460 cm$^{-1}$ and $\Delta =$ 379, 347, and 234 cm$^{-1}$ for UD77, UD94, and OV90, respectively, we find reasonable agreement between the calculation and the experiment (Fig. \ref{fig:four}a).

Figure \ref{fig:four}c-d presents the outcome of this analysis, along with empirical estimates of other characteristic energies as a function of doping. Based on $\Delta\chi^{\prime\prime}$ between 10 K and 300 K (Fig.~\ref{fig:four}b), we identify four energies, from low to high, as summarized in Fig.~\ref{fig:four}c: (1) the onset of the low-energy depletion (where $\Delta\chi^{\prime\prime}$ crosses zero), $\omega_\mathrm{gap}$; (2) the center of the pairing peak, $\omega_\mathrm{pair}$; (3) the center of the two-magnon peak, $\omega_\mathrm{2mag}^\mathrm{peak}$; and (4) the high-energy leading edge of the two-magnon peak (half-maximum position), $\omega_\mathrm{2mag}^\mathrm{edge}$. We make the following observations:

First, while all energies decrease with doping (Fig.~\ref{fig:four}c), $\omega_\mathrm{2mag}^\mathrm{edge}$ varies only slightly and appears to set an upper-bound for $\omega_\mathrm{2mag}^\mathrm{peak}$ in the extrapolation to zero doping. In previous studies \cite{SugaiPRB2003,TassiniPRB2008} the two-magnon peak has been found to soften and broaden with doping, consistent with our result. However, in the undoped limit (which is not accessible in Hg1201 because the AF parent compound is not available) $\omega_\mathrm{2mag}^\mathrm{peak}$ and $\omega_\mathrm{2mag}^\mathrm{edge}$ are typically found in the 2800-4000 cm$^{-1}$ range, larger than our extrapolated values (Fig.~\ref{fig:four}c). (For convenience, in Fig.~\ref{fig:four}c we use the definition of $J^*=\omega_\mathrm{2mag}^\mathrm{peak}/3$, same as in \cite{SugaiPRB2003}, which gives $J^*$ slightly larger than in our model calculation.) We speculate that $\omega_\mathrm{2mag}^\mathrm{edge}$ is related to the bare AF exchange interaction $J$ which shows only weak doping dependence in other cuprates \cite{LeTaconNatPhys2011}.

Second, with increasing doping, both the pairing peak and the pseudogap increase in signal amplitude (Fig.~\ref{fig:four}b), and the values of $\omega_\mathrm{gap}$ and $\omega_\mathrm{pair}$ track each other. This implies that the ERS pseudogap is connected to the pairing peak, even though our data do not allow us to conclusively determine whether they share the same onset temperature. Since the onset temperature of the pairing peak is highest in the most underdoped sample UD77 (Fig.~\ref{fig:three}), this temperature (possibly identical to $T_\mathrm{gap}$) might indicate the mean-field $T_\mathrm{c}$ \cite{DoniachPRB1990,*EmeryNature1995} and be related to the increase in $\omega_\mathrm{gap}$ and $\omega_\mathrm{pair}$ with decreasing doping. The characteristic temperatures $T_\mathrm{gap}$ as defined by the $10\%$ depletion are considerably lower than the pseudogap temperature $T^*$ determined from, \textit{e.g.}, the in-plane resistivity: for doping levels similar to UD94 and UD77, $T^*$ is approximately 185 K \cite{GrbicPRB2009} and 250 K \cite{BarisicPRB2008}, respectively. The difference may be related to the presence of multiple characteristic temperatures above $T_\mathrm{c}$ \cite{GrbicPRB2009,DubrokaPRL2011}, which might further depend on the time scale of the probe.

Finally, we find no clear correlation between $J^*$ and $T_\mathrm{c}$ near optimal doping in a comparison with other compounds (Fig.~\ref{fig:four}d) including La$_{2-x}$Sr$_x$CuO$_4$, which has a relatively low $T_\mathrm{c}^\mathrm{max}<40$~K. All of them have nearly the same $J^*$ for $p\sim0.16$. This implies that other factors affect the attainable $T_\mathrm{c}^\mathrm{max}$, as has been suggested by other authors \cite{TassiniPRB2008,HueckerPRB2011}.

To conclude, we have observed a correlation among the temperature dependences of the two-magnon peak, the pseudogap, and the pairing peak in a model cuprate high-$T_\mathrm{c}$ superconductor. In the overdoped regime, this correlation can be attributed to a feedback effect of Cooper pairing on high-energy magnetic excitations, analogous to the low-energy resonant mode observed by INS \cite{Sidis2004,*YuPRB2010}. This is consistent with anomalies observed in various fermionic spectral functions \cite{Kordyuk2006,Hwang2007,*vanHeumenPRB2009} and directly supports prior indications of a substantial contribution of high-energy magnetic fluctuations to the pairing interaction \cite{DahmNatPhys2009,LeTaconNatPhys2011}. The observation of a closely similar feedback effect in the pseudogap regime is consistent with prior reports of superconducting correlations above $T_\mathrm{c}$ \cite{GrbicPRB2009,DubrokaPRL2011,CorsonNature1999,*BergealNatPhys2008,*YangNature2008,*LeeScience2009,*WenPRL2009,*LiPRB2010},
although other ordering phenomena \cite{Bourges2011,*DaouNature2010,*CaplanPRL2010} and excitations \cite{LiNature2010} may also contribute to this effect in the underdoped samples.

\begin{acknowledgments}
We wish to thank P. Bourges, D.J. Scalapino and C.M. Varma for valuable discussions, and A. Schulz for technical assistance. Y.L. acknowledges support from the Alexander von Humboldt foundation.
\end{acknowledgments}

\bibliography{Hg_B1g_submission}

\end{document}